# Challenges and Opportunities for Bioenergy in Europe: National Deployment, Policy Support, and Possible Future Roles

F. Wu[1,2,*], S. Pfenninger[2]

[1] Climate Policy Group, Institute for Environmental Decisions, ETH Zürich, 8006, Zürich, Switzerland.

[2] Faculty of Technology, Policy, and Management (TPM), Delft University of Technology, 2628 BX, Delft, The Netherlands.

[*] Corresponding author: fei.wu@usys.ethz.ch

**ABSTRACT**

Bioenergy is currently a major renewable energy source in Europe but faces an unclear future because of conflicting modelling results and the lack of long-term policy. This paper identifies three challenges and potential opportunities by analysing bioenergy's historical national deployment, current policy support, and possible future roles in Europe. The first challenge is on the supply side. Calculating the supply-consumption dynamics and import dependency of EU bioenergy, we find that the security of bioenergy supply is challenging for liquid biofuels and those countries with the highest per-capita bioenergy consumption in Europe. Second, the definition of "sustainable bioenergy" in modelling studies is sometimes inconsistent with how EU policies label it. Third, on the demand side, there are unique but competing uses for bioenergy without a clear long-term strategy in Europe. We conclude with three opportunities to tackle these challenges for future research. First, utilising the untapped bioenergy potential with low environmental impacts could improve supply security. A clear and harmonised definition of "sustainable bioenergy" could better convey modelling results to policymaking. Finally, understanding where best to use limited sustainable bioenergy supply through sector-coupled energy system models can provide direction for a clearer EU bioenergy strategy towards 2050.

# 1. Introduction

Bioenergy, that is, the use of biomass feedstocks to supply energy, has become a growing renewable energy source in Europe. It is used not only in heating and cooling (increasing from 66% to 90% of the total renewable heat from 1990 to 2018) but also for bio-blending transportation fuels and subsidised bioelectricity [1,2]. On the one hand, bioenergy is foreseen to play a role in the future EU energy system, including several 100% renewables scenarios in 2050 and the archetypal scenarios of IPCC SR1.5 [3–5]. Recent studies have also argued that biomass could supply "the final few percent" of renewable electricity, enhancing the short-term supply-side flexibility thanks to its dispatchable ability [6,7]. Meanwhile, EU policy envisioned it to play a role in increasing EU energy self-sufficiency [8]. With the recent surging natural gas prices in Europe [9], domestic biogas could be a potential substitute with higher resilience to international trade shocks.

On the other hand, bioenergy is hotly contested because of questionable emissions neutrality – its embodied emissions might be higher than those of fossil fuels when converting carbon sink into energy crops fields [10]. In addition, dedicated biomass can cause potential land-use conflicts between agriculture, forestry, and ecosystem restoration [11–13]. Hence, the contentious bioenergy has been excluded from some of the widely-cited energy systems modelling studies [14–16]. The latest EU Renewable Energy Directive II certification scheme introduces stricter sustainability criteria on all kinds of biomass by 2030 [17]. Several European countries are also phasing out some national support for bioenergy (e.g., The Netherlands stopped subsidising biomass-fired power stations, and Switzerland banned transportation biofuel from mineral oil tax exemption as of 2020 [18,19]). Naylor and Higgins [20] argued that the rapid development of biodiesel would not have occurred without strong policy support, agricultural subsidies, and trade policies. Thus, the phased-out subsidies and stricter sustainability criteria may reduce bioenergy use in the near term, especially in transportation.

All these interdisciplinary concerns over sustainability, policy, and energy systems planning are tangled together, adding to the unclear future of EU bioenergy. While there is a general expectation that bioenergy will continue to play some role in the EU's energy system, it is unclear whether that role will be a system-critical one and whether it can fulfil that role in an environmentally sustainable manner. Current literature has generally explored the role of bioenergy from only one of these concerns (e.g., political frameworks [1], economics and markets [21,22], techno-economic modelling [23,24], or the environment and economics [25]). However, few studies have investigated the interaction between two or more concerns providing more comprehensive and interdisciplinary insights. Moreover, the challenges and opportunities of EU bioenergy development may lie beyond these isolated aspects, shaped by interactions among historical deployment, current policy, and possibilities identified in modelling studies.

This paper intends to provide a more holistic picture of the status quo and potential of bioenergy in Europe, with a focus on these interactions. We consider not only the whole European region but also the national heterogeneity whenever necessary and possible. Although previous studies mostly investigated bioenergy deployment either on the whole EU level [23] or on single countries [26–28], an EU-wide national breakdown is necessary. Because natural resource endowments, historical energy system structures, policies, and support schemes vary across member states. Such national heterogeneity may manifest different challenges and opportunities for bioenergy in Europe, hence providing local policymakers with varied implications.

Here we perform such an assessment to examine EU bioenergy from three areas – first, the spatiotemporal trends in terms of supply-consumption dynamics and energy security; second, EU level policies and national support schemes; third, unique roles for bioenergy discussed in the literature and modelled in decarbonisation scenarios for 2050. We conclude by synthesising challenges and opportunities for bioenergy in a fully renewable and sustainable European energy system.

## 2. Data and Methods

Here we adopt the International Energy Agency's (IEA) classification and dataset of biofuels and waste for energy – that is, world energy balance datasets, one widely-used database providing authoritative and up-to-date renewable energy statistics every year [2]. It includes major biofuels and waste for bioenergy, as shown in Figure 1, of which products are the finest level of available bioenergy data (e.g., solid biofuels, biogases, charcoal, industrial waste, etc.). Note that we do not intend to provide a detailed classification of all possible biofuels/biomass but to include major carriers with consistent European statistics as much as possible.

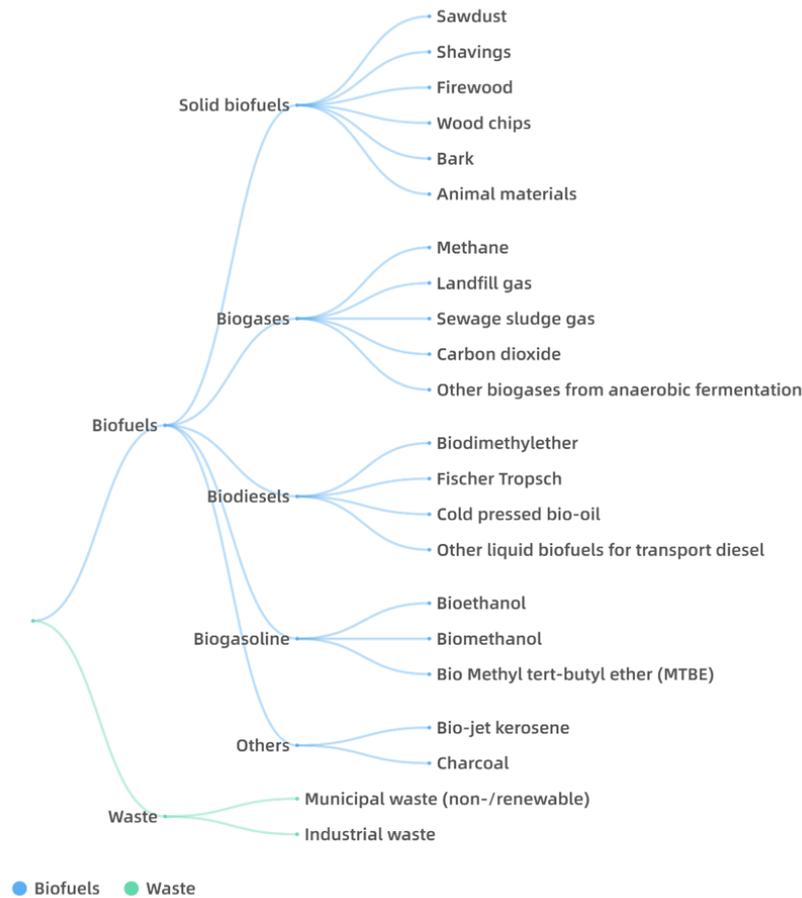

Figure 1: Classification of major bioenergy products and feedstock adopted from the IEA *[29]*

For the studied region and time span, we consider the EU-27 and the UK, excluding their overseas regions, and the timeframe from 2000 to 2018 (including 2019 and 2020 when available). 2018 is the latest year with complete records of energy balance tables from the International Energy Agency (IEA) [2], which is our primary data source for exploring the spatiotemporal trends at the national level.

To analyse bioenergy policies and calculate national support schemes (Table 4), we draw on mixed qualitative and quantitative data sources. They include RES LEGAL Europe database [30] for feed-in and premium tariffs, and the International Energy Agency Bioenergy's national annual reports [18] for bioenergy support levels and biofuel blending quotas. To calculate the shares of national subsidised bioelectricity (Table 2 and Figure 7), we extract subsidy data from the Status Review of Renewable Support Schemes in Europe from 2009 to 2017 by the Council of European Energy

Regulators (CEER) [31], which is the only available open data on bioenergy subsidies. Considering the data availability of European bioenergy support schemes, we adopt the four kinds of support schemes from the CEER [31], including feed-in tariffs (FIT), feed-in premiums (FIP, sometimes also premium tariff, PT), green certificates (GC), and investment grants. A feed-in tariff is a fixed-price design regulating electricity prices through a given amount of per kWh payment to the generators depending on different technologies (irrespective of the wholesale prices), while a feed-in premium adds a bonus to the wholesale market price received by producers [32]. Green certificates are tradable commodities generated with certain renewable electricity providers and may have minimum prices.

Modelling-wise, we select three distinct Shared Socioeconomic Pathways (SSPs) scenarios to compare how varied transition pathways would alter the roles for bioenergy within Europe. Specifically, we compare the sustainability, middle-of-the-road, and fossil-fuelled development SSPs and how their bioenergy supply and total primary energy supply in Europe vary in 2050. The modelling results are extracted from the IAMC 1.5°C Scenario Explorer 2.0 [33] coming primarily from established integrated assessment models. We use the Shared Socioeconomic Pathways (SSPs) here also because they are among the few 2050 scenarios providing specific bioenergy data and Europe-wide results. All the data and the code for processing, analysis, and visualisations, are available on GitHub (see Data availability section).

## 3. Results

### 3.1. Historical deployment of bioenergy

#### 3.1.1. Supply-consumption dynamics and import dependency.

EU bioenergy has contrary trends of increasing total supply volume and decreasing shares in total renewable supply since 2011 (Figure 2). Meanwhile, total bioenergy final consumption has been stabilising, thus creating a growing "gap" between its total energy supply and total final consumption. This "gap" is primarily the bioenergy used for "Transformation processes" – converting primary biomass feedstocks into secondary/intermediate energy carriers (e.g., intermediate heat from CHP plants). In 2018, the transformation processes at heat, electricity, or CHP plants in Europe used over 2700 TJ bioenergy, which are subtracted from the "total final consumption" in the world energy balances databases [2].

Further broken down by product, primary solid biofuels have remained the dominant source of bioenergy in all EU countries with a growing supply volume but with a falling share of overall bioenergy use (from 81% of all bioenergy in 2000 to 60% in 2017). Other products have growing shares, with biogas rising from 3% to 10% and biodiesel from 1% to 11% during the studied period. The waste sector was minor and stable, constituting around 15% of the total supply, with a steadily decreasing share of industrial waste from 3.5% to 2.3%. Though dominated by primary solid biofuels, the EU's bioenergy supply structure has become more diversified. The majority of biofuels are domestically produced, while the rise of biodiesel has been primarily driven by imports, of which over one-third are from outside the EU (Figure 3). The EU has shifted from a domestic bioeconomy in 2000 to one heavily depending on sourcing liquid biofuels from overseas. The import dependency for biodiesels and biogasoline has risen from close to 0% to over 60% in less than 20 years (Figure 3), though these are primarily used for blending with and thus displacing imported fossil transport fuels.

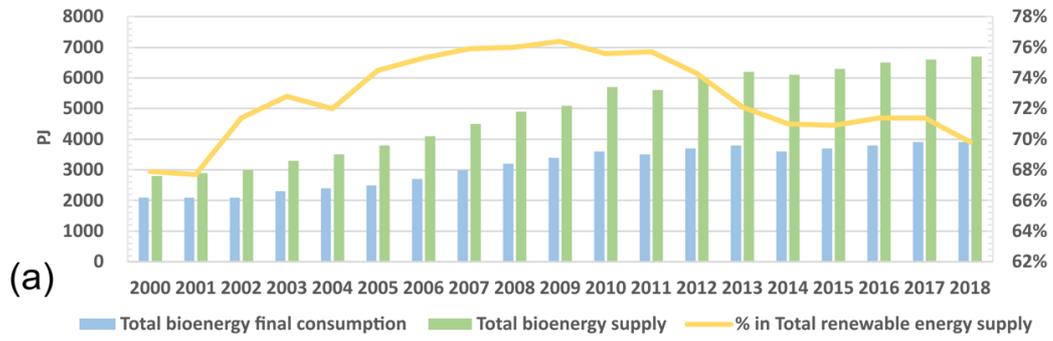

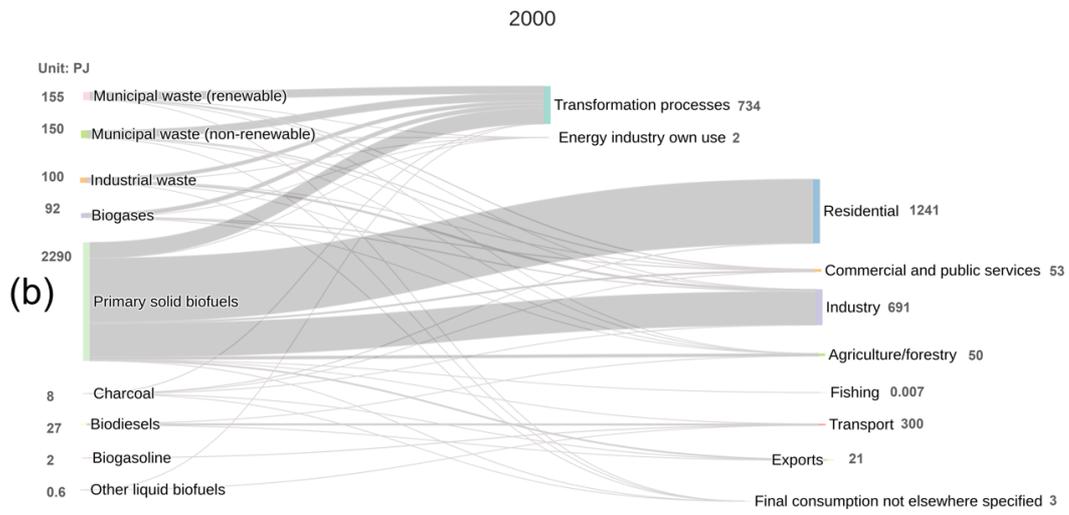

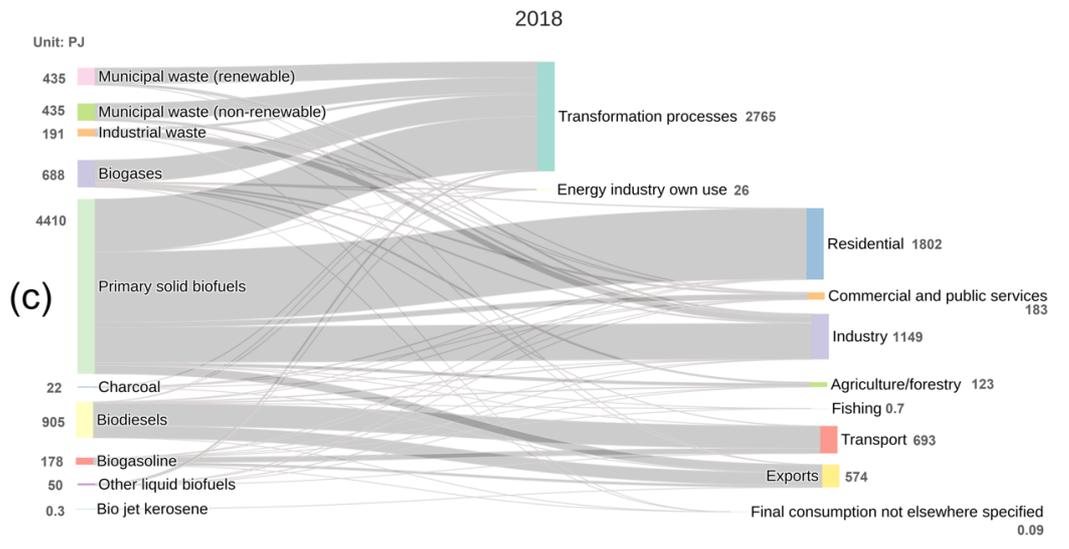

Figure 2. Overall EU bioenergy deployment. (a) Trends of EU bioenergy total consumption, supply, and its shares in renewables. (b) 2000 Energy flows of bioenergy supply, intermediates, and end-use sectors. (c) 2018 Energy flows of bioenergy supply, intermediates, and end-use sectors. 2018 is the last year with complete data records. Snakey diagrams (b) and (c) are in the same unit (PJ). To enable direct comparison between (b) 2000 and (c) 2018,

the width of the colored bars in the Sankey diagrams are in proportion to the energy flow.

On the consumption side, mirroring the rise of biogas and biodiesel, the share of residential (i.e., households) consumption dropped from 58% to 41% within this period. Driven by the legally binding EU-wide target of 10% renewable energy used for transportation (or even higher at the national scale, e.g., 20% in Finland), biofuels have had growing importance for the transportation sector, with the share of transportation in consumption dramatically rising from 1.5% to 14.7% by 2017. However, as most of the liquid fuels for transportation are imported – about 97% of the crude was imported in the EU in 2019 [2] – transport biofuels currently replace one source of import dependency (imported fossil fuels) with another (imported biofuels). The potential of the clean energy transition to reduce import dependency thus remains unaddressed (Section 3.2).

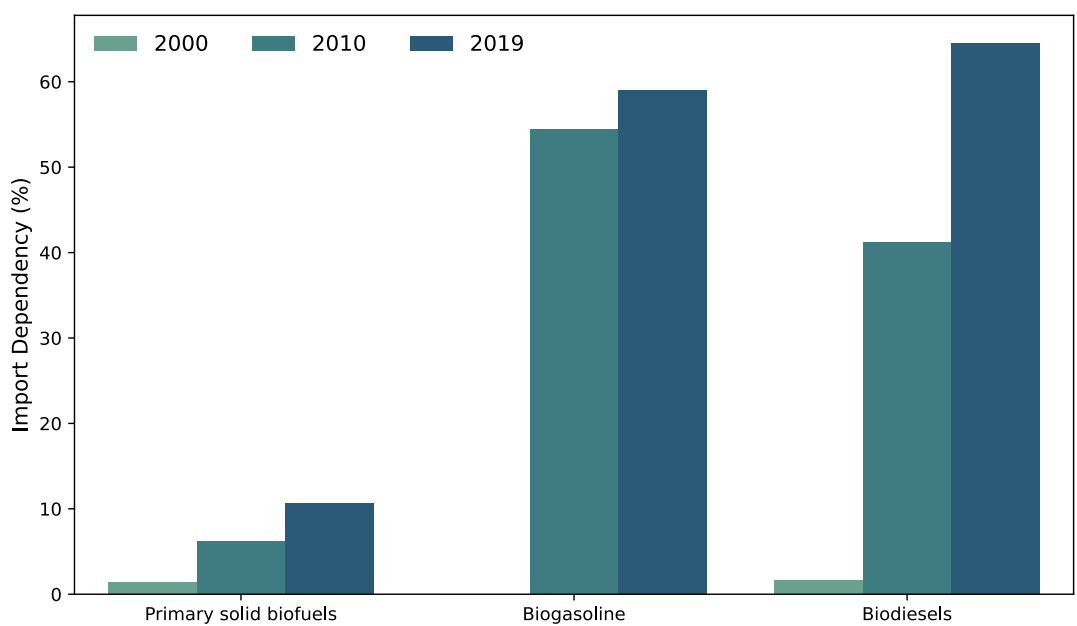

Figure 3. Changes in import dependency for the main biofuels (import dependency = imports/total energy supply per biofuel type).

3.1.2. Per-capita differences across countries.

Sweden is notable for the highest proportion of bioenergy used for industry. In absolute terms, France and Sweden were the largest bioenergy supply countries in 2000, but were surpassed by Germany by 2017. However, considering the large population in Germany, bioenergy consumption per capita presents a rather different picture. As shown in Figure 4, the biomass used for power, heating, and transport per capita differs substantially across EU countries.

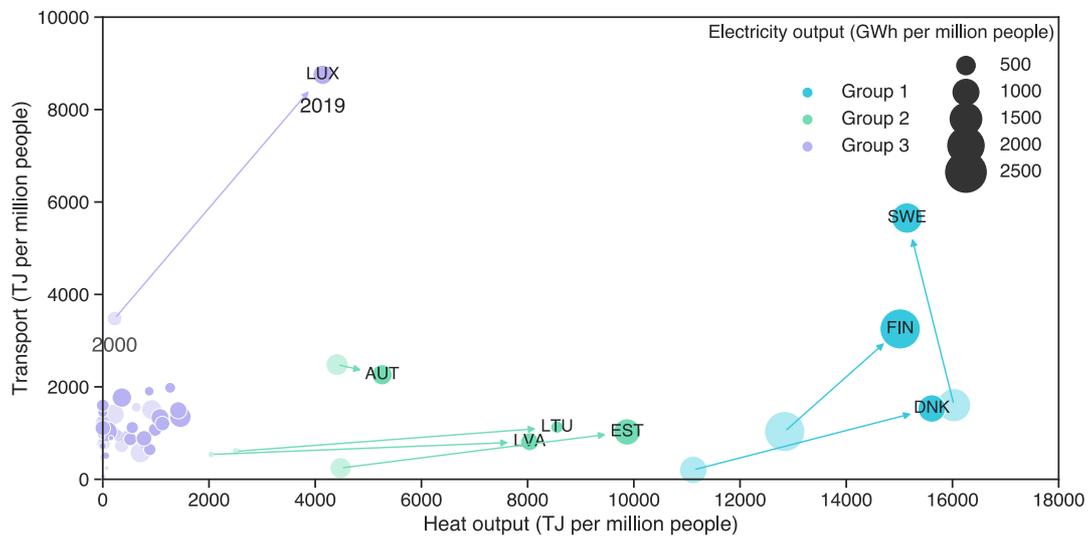

Figure 4. Changes in bioenergy per-capita consumption (transport, heat output, and electricity output per million people) from 2000 to 2019.

Three groups of nations stand out, of which the leading group consists of Finland, Sweden, and Denmark (Group 1). This group is characterised by high per-capita levels of bioenergy consumption (especially for electricity and heat) throughout the whole period. Though the per-capita levels of heat output in group 1 countries initially varied, they have converged by 2019. Group 2 consists of Estonia, Latvia, Austria, and Lithuania with higher per-capita heat from biomass. They start within the general mass of EU countries in 2000 but move towards Group 1 countries by 2019.

The remaining EU countries constitute Group 3, showing an overall low level of per-capita bioenergy consumption in all three sectors. However, one outlier is Luxemburg, where the per-capita biofuels for transportation have soared – it is now the number one country in the EU on that metric, despite its merely average biomass use for electricity and heat. This unique phenomenon is partly because the international work commuters contribute to transportation demand but not to the population or residential statistics.

## 3.2. Policy and support schemes for bioenergy

### 3.2.1. EU policy: stricter sustainability criteria yet unclear role of bioenergy.

Overall, most bioenergy-related EU policies focus on the stricter sustainability criteria of bioenergy and mandate its growing targeted share in the transport sector or together with other renewables in the gross final energy consumption. Considering these two aspects, we list the major policies in chronological order along with the changing share of bioenergy products (Figure 5) and then compare these policies in Table 1.

In 2003, the European Commission (EC) issued the EU Directive on Biofuels, focusing on a first blending target (5.75% of biofuels by 2010) for the transportation sector [34]. Furthermore, to encourage the widespread use of bioenergy not only for transport but also for heating and electricity, the Biomass Action Plan (European Commission, 2005) first emphasised the importance of the bioenergy industry [35], specifying the general sustainability criteria of biofuels, including GHG reduction and biodiversity (Table 1). The EU strategy for biofuels issued in the following year further complements the Biomass Action Plan by a threefold objective with seven strategic policy areas –"further promotion of biofuels in the EU and developing countries, preparation for the large-scale use of biofuels, and heightened cooperation with developing countries in the sustainable production of biofuels" [36].

These earlier policies paved the way for the EU Renewable Energy Directive (RED-1) issued in 2009. RED-1 complemented the former plans with overall mandated goals, i.e., 20% renewables in the gross final energy consumption and 10% renewables in transport by 2020. This latter target is essentially met by transport biofuels alone [37]. The directive further committed every country to set their annual breakdown of bioenergy shares by 2020 in the National Renewable Energy Action Plans (NREAPs) [38]. However, the RED-1 did not include life-cycle GHG emissions of biofuels caused by indirect land-use change, which could be even higher than those of fossil fuels when natural ecosystems with higher carbon stock are converted into agricultural land for energy crops [10]. Recognising this critical missing concern, the European Commission amended the RED-1 in 2012 by a legislative proposal that includes guidelines to estimate indirect land-use change emissions from biofuels (ANNEX V) [39] as well as capping conventional biofuels and promoting advanced biofuels [37,40].

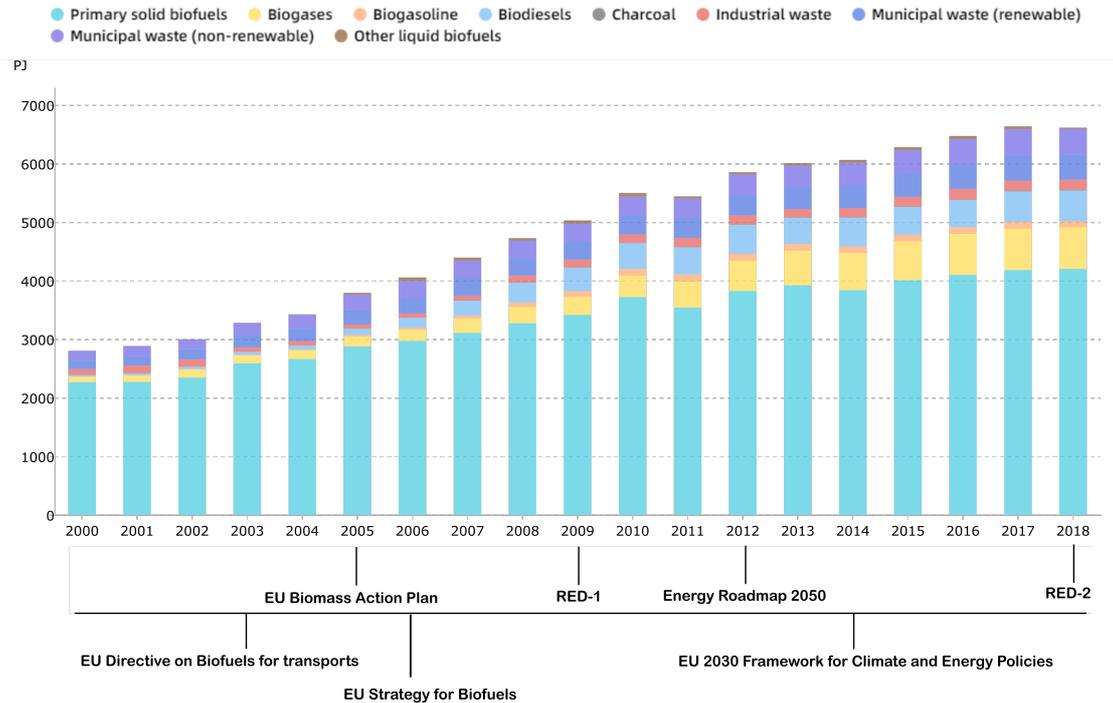

Figure 5. Timeline of major policies and the share of bioenergy products in the EU from 2000 to 2018.

The Energy Roadmap 2050 [41] and EU 2030 Framework for Climate and Energy Policies [42] further proposed to increase the overall EU target of renewable shares and GHG reduction. However, no specific national bioenergy target has been suggested beyond 2020 or towards 2050. Most recently, the EU Renewable Energy Directive II recast (RED-2) sets an increased target of 32% renewables in the gross final energy consumption, along with a minimum goal of 14% renewables in transportation by 2050 (advanced biofuels double count and should reach 3.5%) [43,44]. Furthermore, the sustainability framework of bioenergy is reinforced in RED-2 with (1) detailed GHG criteria and calculation rules for solid biofuels and biogas, (2) new sustainability criteria for forestry biomass, as well as (3) a new approach limiting biofuels with high indirect-land-use-change risk. For instance, palm oil, a traditional source of biodiesel, has more than 40% expansion on high carbon stock land, thus classified as a high indirect-land-use-change-risk biofuel feedstock [45]. Therefore, it is likely that the supply of biofuels, especially the traditional biofuels from energy crops and forests, will see a decrease in response to the EU's stricter sustainability criteria.

Contrary to RED-1, RED-2 generally commits nations to establish support schemes for expanding renewables in an "open, transparent, competitive, non-discriminatory, and cost-effective" way [43], which does not specify what should happen with regards to bioenergy. The NREAPs and RED-1 terminated in 2021, and the new RED-2 is to come into force with no national binding targets or specific breakdown of bioenergy; there seem to be no clear policy targets on further bioenergy deployment to meet the EU and member states' 2030 ambition and a 100% renewable Europe in 2050.

Table 1. Development of major bioenergy-related policies and directives in the EU.

| Policy/Directive | Year | Targets (if applicable) | Sustainability Criteria |
| --- | --- | --- | --- |
| EU Directive on Biofuels for Transport [34] | 2003 | 5.75% of biofuels in transport by 2010 | Not specified |
| EU Biomass Action Plan [36] | 2005 | Not specified | Saving 35% GHG emissions compared to fossil fuels; Cannot be sourced from areas with high carbon stock & biodiversity |
| EU Strategy for Biofuels [46] | 2006 | Prioritising the role of biofuels in transportation by a threefold objective with seven policies areas | Not specified |
| EU Renewable Energy Directive (RED-1) [38] | 2009 | 20% renewables in the EU gross final energy consumption and 10% in transport by 2020 (Breakdown of bioenergy stipulated by every country in NREAPs) | Rising to 50% of GHG savings and 60% for new plants from 2018 and onwards [39] |
| Energy Roadmap 2050 [41] | 2012 | Bioenergy should contribute 22–28% of the EU gross inland energy consumption in 2050 | Not specified |
| EU 2030 Framework for Climate and Energy Policies [42] | 2014 | Not specified | A collective delivery and commitment to a 40% reduction in GHG emissions by 2030 |
| EU Renewable Energy Directive II recast (RED-2) [43] | 2018 | Increased shares of renewables to 32% and a minimum of 14% within the transport (3.5% of advanced biofuels) by 2030 | Reinforced sustainability framework: 65% of GHG savings for transport biofuels; New criteria for forestry feedstock and GHG criteria for solid and gaseous biomass fuels; A new approach limiting biofuels with high indirect-land-use-change risk [47] |

3.2.2. Low national subsidies in high per-capita bioenergy countries.

Motivated by the collective EU targets, member states have set up different national support schemes for bioenergy used in power and heating sectors and mandated biofuel blending targets for transport. For the power sector, feed-in tariffs and premium tariffs are the primary support schemes encouraging biomass for electricity (Figure 6 and Table 4). Notably, the leading countries in terms of per capita bioenergy consumption (Group 1 identified in section 3.1) had low levels of unit support, which were 19.49 €/MWh (Finland), 12.80 €/MWh (Sweden), and 35.89 €/MWh (Denmark) in 2017 – they have been utilising the highest per capita level of bioenergy at the cost of low unit supports.

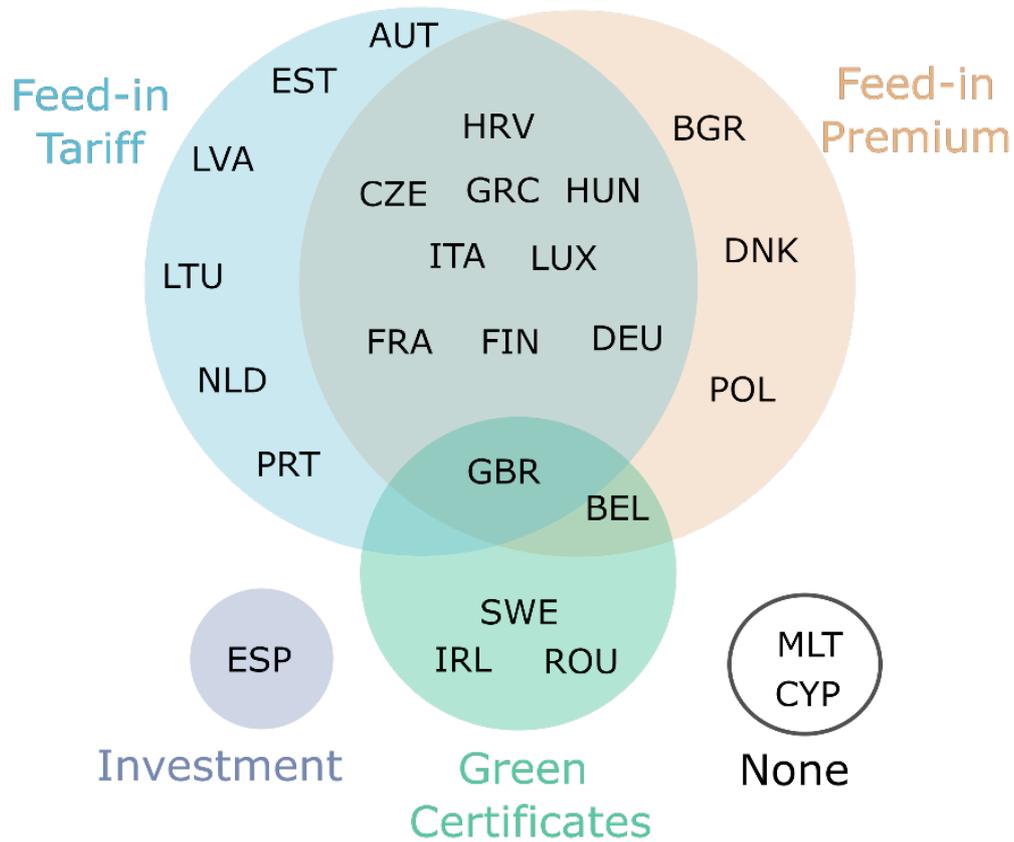

Figure 6. Support schemes for bioelectricity across the EU (2017).

As a consequence of this variety of supporting schemes, the subsidised bioelectricity share ranges from 2% (France, Lithuania, Portugal, and Romania) to over 30% (Ireland, UK) in 2017 (Figure 5 and Table 3). Overall, most member states have displayed an increasing trend of bioelectricity subsidies, with Ireland (35%), the UK (33%), and Germany (24%) having the highest shares of supported bioelectricity output in 2017, indicating the dominant role of support schemes for fostering the use of biomass in the power sector in these countries. Meanwhile, subsidies in nine countries declined, including Austria, Cyprus, Spain, Estonia, France, the Netherlands, Poland, Portugal, and Sweden. Sweden, Portugal, and Estonia, in particular, have seen a sharp drop in supported share by around half in the studied period. Sweden has one of the highest per-capita consumptions of bioenergy in the EU, yet with low levels of financial support – both the shares of subsidised electricity and unit support cost are low.

Table 2. Shares of subsidised bioelectricity in the EU. ("-" represents unavailable data. Author's calculation by compiling the annual Status Review of Renewable Support Schemes in Europe [31]. 2012 is the earliest year, and 2017 is the latest update). 1 Group 1 and 2 Group 2 are countries clusters identified in section 3.1.

| Country | 2012 | 2017 | Country | 2012 | 2017 |
|---------|------|------|---------|------|------|
| AUT[2] | 4.00% | 3.62% | HUN | 2.61% | 11.80% |
| BEL | 12.00% | - | IRL | 10.58% | 34.61% |
| CYP | - | 11.47% | ITA | 10.23% | 18.65% |
| CZE | 5.46% | 8.76% | LTU[2] | 1.51% | 2.21% |
| DEU | 11.16% | 23.56% | LUX | - | 13.33% |
| DNK[1] | 11.67% | 14.68% | LVA[2] | - | 3.06% |
| ESP | 6.93% | 5.21% | NLD | 10.69% | 9.59% |
| EST[2] | 7.41% | 4.43% | POL | 8.85% | 4.52% |
| FIN[1] | 2.07% | 4.18% | PRT | 3.71% | 2.71% |
| FRA | 2.86% | 2.27% | ROU | 0.34% | 2.02% |
| GBR | 17.09% | 33.36% | SVN | - | 3.37% |
| GRC | 1.38% | 2.28% | SWE[1] | 8.68% | 4.00% |
| HRV | 0.42% | 3.95% | | | |

Though there are fewer direct support schemes for biomass used in the heating sector, it may receive subsidies for CHP (combined heat and power) plants jointly with bioelectricity or gain connection priority in some countries [31]. In addition to support schemes for CHP and district heating, other instruments for energy recovered from waste are also used. For Sweden, the taxation and charge for energy recovery from residual waste are increasing (e.g., SEK 75/tonne in 2020, SEK 100/tonne in 2021, to increase further) [48]. As a result, however, the country recovers more energy from waste than any other European nation (3 MWh/tonne in 2019) [48].

Unlike in the power and heating sectors, biofuels for transportation have been indirectly mandated by the EU through the minimum of 10% renewables consumed by the transport sector in every Member State in 2020. Correspondingly, countries have adopted different minimum mandates for biofuel quotas (Table 3), which is another reason why the end-use of bioenergy has been soaring in the transport sector, apart from the general EU-wide target. The biofuels quota may slightly vary depending on the blending fuel type (e.g., 5% for E5, 10% for E10, and 7% for diesel in Belgium). Although most EU countries set blending targets around 10%, Finland has the highest goals of 20% by 2020 and 30% by 2030. As transport has been the second-largest $CO_2$ emitter in Finland (2016), this could play a role in helping reduce $CO_2$ emission outside the EU Emissions Trading Scheme (ETS) with cost-effective biofuels supply and ambitious quota obligations – but whether transport fuel blending is the most useful use of limited biofuel is nevertheless open question, to which we want to start turning our attention next.

Table 3. Levels of bioenergy support schemes in different sectors in the EU.

| Country | **Feed-in tariff for electricity** | **Premium tariff** | **Unit support level in 2017** | **Biofuel quota for transportation**[1] |
|---|---|---|---|---|

|  | (€ct per kWh) | for electricity (€ct per kWh) | (€ct per kWh) |  |
| --- | --- | --- | --- | --- |
| AUT | 4.66 – 21.78 |  | 10.80 | 8.45% |
| BEL |  |  |  | 5.00 – 10.00% |
| BGR |  |  |  | 9.00% |
| HRV |  |  | 9.95 | 10.05% |
| CYP |  |  | 13.70 |  |
| CZE | 5.60 – 13.00 | 7.20 – 11.50 | 10.50 | 4.10 – 6.00% |
| DNK |  | 3.50 – 5.30 | 3.59 | 5.75% |
| EST |  |  | 2.10 | 5.00% |
| FIN |  | 8.35 – 13.35 | 1.95 | 20.00% |
| FRA | 12.00 – 17.50 |  | 10.10 | 8.00 – 10.00% |
| DEU | 5.71 – 13.32 | 5.66 – 23.14 | 11.83 | 6.30% |
| GRC | 14.80 – 19.80 | 1.29 – 2.25 | 7.85 | 7.00% |
| HUN | 4.04 – 11.05 | 9.89 | 2.13 | 6.4% |
| IRL | 8.96 – 14.70 |  | 5.37 | 11.11% |
| ITA |  |  | 15.40 | 6.50% |
| LVA | | | 12.28 | |
| LTU | | 1.11 – 1.34 | 5.66 | |
| LUX | 11.70 – 16.20 | 1.51 – 1.90 | 10.10 | 5.85% |
| MLT | | | 10.80 | 10.00% |
| NLD | | 4.6 – 9.2 | | 7.75% |
| POL | | | | 8.50% |
| PRT | 10.20 – 11.90 | | 9.95 | 10.00% |
| ROU | | | 13.70 | 6.50% |
| SVK | 7.03 – 9.22 | | 10.50 | 8.20% |
| SVN | | | 3.59 | 7.50% |
| ESP | | | 2.10 | 8.50% |
| SWE | | | 1.95 | 13.80%[2] |
| GBR | 1.97 – 5.58 | | 10.10 | 10.64% |

[1] For countries with different biofuel quotas throughout the time, the latest one has been listed here if available (i.e., the target for 2020)

[2] Sweden no longer divides the renewable energy target into further targets per sector, so the projection in the NREAP is displayed instead.

### 3.3. Possible roles of bioenergy from modelling studies

#### 3.3.1. Varied modelling assumptions of sustainable bioenergy supply.

On the supply side, both the future potential and share of bioenergy varies among scenarios due to inconsistent modelling assumptions of "sustainable bioenergy". But one common consistency is that the stricter sustainability, the lower bioenergy potential/deployment. For instance, the Shared Socioeconomic Pathways scenarios present three future narratives – SSP1 (sustainable development with well-managed land systems and limited societal acceptability for Bioenergy with Carbon Capture and Storage, BECCS), SSP2 (a middle-of-the-road scenario where it follows historical societal and technological development), SSP5 (fossil-fuelled development with intensive resource and energy consumption mitigated through BECCS). Comparing the three distinct pathways from six models (Figure 8), the shares of bioenergy in primary energy supply relate to whether carbon capture and storage is restricted (SSP1) or intensively deployed (SSP5).

Even when comparing the same scenario, bioenergy potential still varies significantly among models (e.g., SSP1 in Figure 8). This is due to various modelling assumptions and definitions of what is "sustainable bioenergy" (Table 2). One major difference is how modellers consider different biomass feedstock types and their sustainable potential. Currently, most models have included dedicated energy crops or short-rotated forests [49,50]. However, land-use change and environmental sustainability remain the key issues when sourcing bioenergy from dedicated conventional energy crops [11,12]. Therefore, some models prevent such concerns by excluding the dedicated energy crops and using waste and residues only [51,52], though the biomass supply may be limited and insufficient.

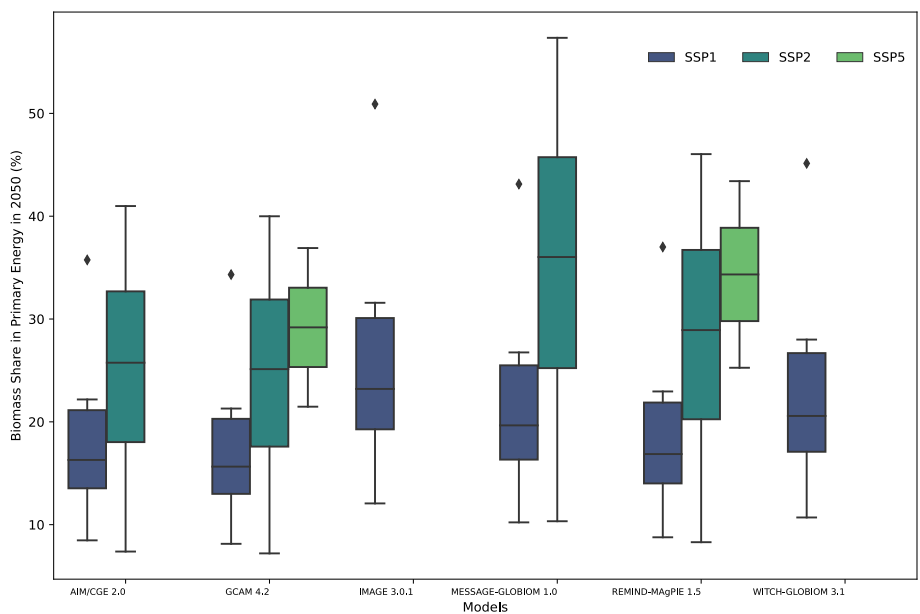

Figure 7. Different shares of biomass in primary energy in 2050 SSP (Shared Socioeconomic Pathways; SSP1 refers to a sustainability scenario, SSP2 is a middle-of-the-road one, while SSP5 depicts the fossil-fuelled development).

Table 4. Different assumptions of biomass feedstock and bioenergy potential in 2050.

| Models (scenarios) | European bioenergy potential in 2050 (TWh) | Biomass feedstock considered |
|---|---|---|
| JRC-TIMES (Low or high availability scenarios) | 2400–5869 | Biofuel crops, dedicated perennial crops, residues from agriculture, forests, and waste [49] |
| PRIMES | 1837 | Biomass and waste [53] |
| IMAGE (SSP2) | 3709–5775 | Biomass from agriculture and forests (i.e., maise, |

|  |  | sugar cane, switchgrass and miscanthus) [50] |
|---|---|---|
| Euro-Calliope | 2400 | Wastes and residues from JRC-TIMES [52] |
| PyPSA-Eur-Sec | 2400 | Wastes and residues from JRC-TIMES [51] |

### 3.3.2. Unique roles of bioenergy with competing end-uses.

Bioenergy may play unique roles that other renewables cannot fulfil, such as balancing intermittency [54–56], providing fuels for hard-to-decarbonise sectors [57], allowing negative emissions [58,59], and enhancing national energy diversity [8]. When the European power system gets close to 100% renewables, bioenergy could help balance fluctuations in renewable power systems otherwise dominantly supplied by weather-driven wind, solar, and hydropower [60] [54]. Technical options could be larger storage capacities of intermediate biomass, combined heat and power (CHP) biomass plants, or biogas upgrading to gas grids, e.g., bio-methane [60]. However, the same biofuel could be used for other competing applications apart from power systems: e.g., methane is also a critical industry feedstock that is hard to decarbonise.

From the negative-emissions perspective, bioenergy is the only renewable source capable of carbon-negative power stations, making it a compelling component of energy systems transition otherwise primarily dominated by wind and solar power. Moreover, with carbon capture and storage technologies, BECCS (bioenergy with carbon capture and storage) is considered a "saviour" of feasibility for most explorative 1.5°C and 2°C climate mitigation pathways [58,61]. This also resonates with the SSPs scenarios in Figure 8, where extensive carbon capture and storage deployment demand the highest bioenergy share. Nevertheless, if we consider the constraints of environmental sustainability, especially land-use impacts, the story may be very different. For example, the biomass supply suggested by 1.5°C pathways would require additional land-use change causing net losses of carbon from the land and overuse of freshwater [58,62].

### 3.3.3. Substantial potentials from ancillary bioenergy.

In contrast to the overall increasing trend in biomass for biofuels, the current utilisation of the municipal and industrial waste sector is relatively stable in Europe, constituting around 15% of the total bioenergy supply (Figure 2). However, recent studies suggest a considerable potential from municipal waste, agricultural residues, by-products, and co-products. All these studies provide a circular-economy perspective to reuse non-traditional feedstocks to provide bioenergy (Table 3).

Existing literature has well explored the separate potential of energy recovered from municipal waste, agricultural and forest sources (residues, co-products, and by-products), respectively, but without considering their combined potentials. All of these products share the common feature of recovering energy by reusing/recycling biomass of little or no value, which would otherwise be left to waste. Though the term "waste-to-energy" is well recognised for energy recovered from municipal waste, it cannot stand for the biomass from agricultural co-products or forest by-products, as waste and co-/by-products are different by definition.

To our best knowledge, there is no existing term to represent such non-dedicated bioenergy from the three different sources of human settlement, agriculture, and forests: we thus define it here as "ancillary bioenergy". In contrast, we define those biofuels/biomass which are intentionally and specifically grown for energy utilisation (e.g., soybeans for biodiesel, corn grain for bioethanol, lignocellulose for renewable heat, etc. [63]) as "dedicated bioenergy". Ancillary biomass still has additional sources for domestic bioenergy without food competition or land conflicts, instead reusing resources in a circular-economy way. Some key feedstocks from ancillary bioenergy could be by-/co-products of high energy density that have not been included in major energy models (e.g., nuts shells, animal fats/oil, used cooking oil, etc.).

We provide an overview of different ancillary bioenergy potentials studied in separate papers (Table 6). Most of them look at the separate potential of a sub-category and are mainly based on historical or current spatial data.

Table 5: Overview of ancillary bioenergy potentials for different sources from the literature, along with the methods used to reach the estimation.

| Feedstock | Potentials | Methods/Models | Reference |
| --- | --- | --- | --- |
| Agricultural residues, by-products, and co-products | 18.4 billion tonnes of total potential in the EU28 (Animal ~31%, Vegetable ~44%, Cereal ~22%) | GIS and statistical analysis | [64] |
| Agricultural residues, by-products, and livestock sewage | 820,000 tonnes of feedstock per year can be used in small-scale CHP units to satisfy the thermal and electric demand of 116,000 households and 178,000 families in the Calabria region (Southern Italy) | Statistical estimation | [65] |
| By-products (fish fats) | 20,000 tonnes of fish oil soapstock are available annually in Norway for producing 51.8 GWh bioenergyAnnual potential of 106,000 GJ energy production per annum in Portugal (Marvão) | Statistical estimation | [66] |
| Used cooking oil | 13% of the biodiesel demand could be met by used cooking oil in BrazilPotential of 2,300 PJ/year in EU27 | Statistical estimation | [67] |

# 4. Discussion

Biomass has grown in importance over the past two decades, but it remains a contentious renewable energy source in Europe, with an uncertain future. We identified challenges and potential opportunities for EU bioenergy by reviewing and analysing its historical national deployment (Section 3.1), current political support (Section 3.2), and future modelling studies (Section 3.3), summarised in Table 6.

We identify three cross-cutting issues and opportunities in particular: supply security and untapped bioenergy potential, gaps between sustainability definitions in EU bioenergy policy and in modelling studies, and the question of optimal allocation in view of competing demand for limited resources. We now discuss these three challenges in turn.

Table 6. Summary of challenges in EU bioenergy national deployment, political support, and modelling studies.

| Sections | 3.1 Historical National Deployment | 3.2 Current Policy & Support Schemes | 3.3 Future Modelling Studies |
|---|---|---|---|
| **Supply challenges** | 1a. Surging import dependency of liquid biofuels (Figure 3)<br><br>1b. National heterogeneity (Figure 4 and Figure 8) | 2a. Stricter sustainability criteria & phase-out of biodiesel from palm (Table 1)<br><br>2b. No long-term bioenergy policy (Table 1) | 3a. Varied sustainable potential & assumptions (Table 5 and Figure 7)<br><br>3b. Untapped potential of ancillary bioenergy (Table 5) |
| **Demand challenges** | 1c. Growing consumption during transformation processes (Figure 2) | 2c. Transport as the key mandated sector (Table 3) | 3c. Unique but competing uses (Section 3.3.2) |

## 4.1. Supply security and untapped bioenergy potential

Although the EU initially designed bioenergy as an important alternative fuel for increasing energy diversity and self-sufficiency [8], over 60% of liquid biofuels were imported in 2019 (Figure 3). The surging import dependency of liquid biofuels (1a) primarily displaces the formerly imported fossil fuels to fulfill the mandated blending targets for transportation (2b). Moreover, with the stricter sustainability criteria on all kinds of biomass coming into force (2a), the domestic biomass supply will likely become even scarcer, thus challenging the security of bioenergy supply in Europe.

The supply security also has national heterogeneity, especially in countries with high per-capita bioenergy consumption and low national subsidies (Group 1 and Group 2 countries, as identified in Figure 4). Consolidating the three dimensions of nationwide bioenergy deployment – subsidised bioelectricity, import dependency, and bioenergy share in the total energy supply, Figure 8 now gives us a distinct picture of the three national groups compared to Figure 4. Generally, most EU countries (purple dots as Group 3) are moving towards a lower bioenergy share and higher subsidies (upper-right of the ternary chart). However, the consumption-leading Groups 1&2 have relatively low and decreasing subsidies at the cost of higher import dependency (blue and green dots moving to the bottom). Noticeably, Denmark (from Group 1) imported about 40% of its woody biomass from countries outside the EU in 2018, e.g., Russia (11%), the USA, and Canada (19%) [68]. Therefore, the security of bioenergy supply is not only a common challenge just for imported liquids among the whole EU but also for the leading per-capita consumption countries.

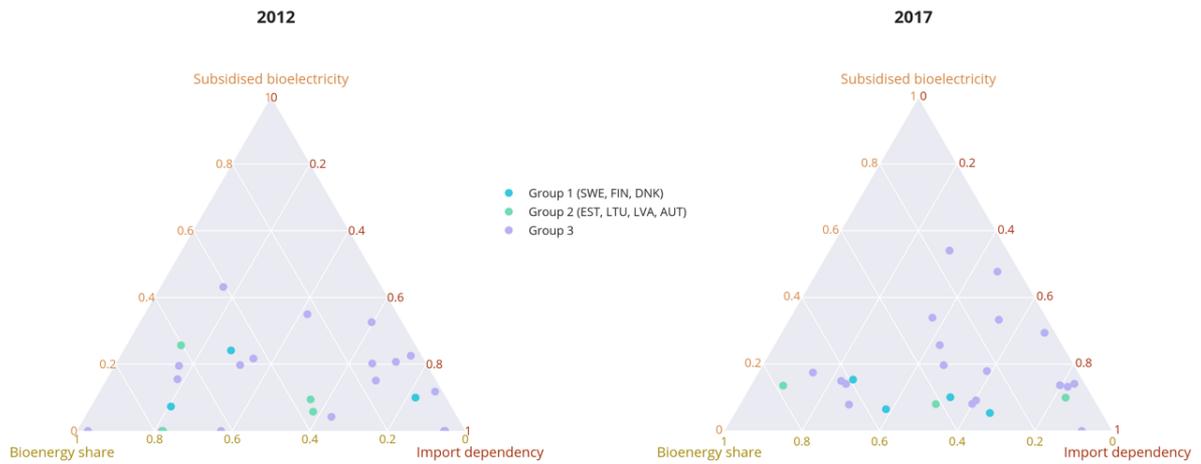

Figure 8. Comparing shares of bioelectricity received subsidies, bioenergy in total energy supply, and its import dependency (For bioenergy subsidy data, only the subsidised bioelectricity is available from 2012 to 2017 *[31]* )

One possible opportunity to combat this challenge could be exploring the "extra" or untapped bioenergy potential. Ancillary bioenergy, as proposed in Section 3.3.3., has substantially untapped potential, which could add to domestic supply without land-use competition. The collective potential of ancillary bioenergy (i.e., non-dedicated bioenergy from human settlement, agriculture, forests, and waste) is not systematically explored in the literature (Table 5)., nor has it been considered in future European energy scenarios so far (Table 4). A systematical estimation of the collective potential of ancillary bioenergy for future energy systems is lacking.

In addition to ancillary bioenergy, abandoned land or dietary shifts could also provide "extra" biomass without food-/land- conflicts [69]. For instance, modelling results suggest that reducing beef, lamb, and other land-intensive food would result in an extra land supply for biofuel crops without conflicting with food production [70]. Moreover, replacing animal-based diets with plant-based ones could achieve a 70% reduction in land use and the associated GHG emissions [71]. However, all the untapped "extra" bioenergy is not a panacea, but a set of alternative futures with pros and cons. Whether it is economically sensible to collect ancillary bioenergy, how would waste reduction influence its availability, and how to sustainably allocate biomass on abandoned land – these research questions require more interdisciplinary modelling studies to provide plausible options for policymakers.

## 4.2. Supply sustainability: Gaps between EU bioenergy policy and modelling studies

Just as how versatile biomass feedstocks are, the definition of sustainable bioenergy or even just bioenergy varies. There is currently no unified classification for bioenergy in the literature. It can be categorised by fuel states (solid, liquid, and gaseous) [72], by sources (e.g., FAO's classification from energy crops, agricultural residues, by-products, municipal waste, etc.), by generation (first-, second-, and third-generation biofuels), or through combined criteria for statistical purposes (e.g., IEA, IPCC). Some harmonised systems for classifying bioenergy and biomass inputs have also been advocated (e.g., [73]), but no consensus has been reached. Similarly, "sustainable bioenergy potentials" are more varied among modelling studies (3a in Table 6). This is a common issue for modelling studies given their different assumptions of "sustainable" and various data sources.

However, this could be especially challenging for EU policymakers due to the gap of "sustainable bioenergy" definition between existing literature and ongoing policies. In other words, what energy modelling studies label "sustainable bioenergy" is not always consistent with how EU policies define it. For example, with the new EU bioenergy sustainability certification scheme (Annex IX in EU Renewable Energy Directive II, RED II [47]), some dedicated energy crops with high indirect land-use risks (e.g., palm oil) will be phased out even if they fulfil previous sustainability requirements [45]. Meanwhile, the indirect land-use change and its embodied emissions are still poorly represented in some widely-cited 2050 modelling scenarios. E.g., the biomass supply suggested by 1.5°C pathways would require additional land-use changes causing net losses of carbon from the land and overuse of freshwater when deploying extensive Bioenergy with Carbon Capture and Storage technologies (BECCS) [58,62]). Since the BECCS is intensively advocated and deployed in most 1.5°C and 2°C climate mitigation pathways [58,61] and the Shared Societal Pathway 5 (Figure 8), whether the sustainable bioenergy supply is sufficient to support these negative emissions is contentious for policy implications.

Rather than adopting existing sustainability definitions from policy, energy modellers could take a step further by investigating higher sustainability bars for modelling future energy systems, especially for ruling out indirect/undesirable land-use change. There is an opportunity here for clearing up the definitions of what "sustainable biomass supply" really is so that energy system studies can determine whether this sustainable biomass supply is sufficient for its intended purpose.

## 4.3. Competing demand and optimal allocation

Demand-wise, there are currently no sector-specific goals or nationally binding targets for bioenergy from 2021 onwards in Europe, except for the blending of transportation fuels. However, biomass has other possibly unique roles in a 100% renewable and zero-emissions energy system that other renewables cannot substitute, such as decarbonising industry and balancing power grids (3c in Table 6). Hence, the use of bioenergy for transportation fuel blending may remove feedstocks from other more strategically relevant uses in a renewables-based clean energy system. Moreover, there is the risk that bioenergy becomes locked-in to uses like fossil fuel blending without an overarching EU bioenergy strategy considering all energy sectors – what are the possible end-uses for different biomass feedstocks; how to optimise the limited sustainable feedstocks; where is bioenergy competitive over other renewable technologies in the long run?

In addition to the lack of EU bioenergy strategy, European countries have been responding differently through varied national support schemes. On the one hand, some sharply reduce the subsidies for bioelectricity (the highest reduction is over 50% in Sweden, see Table S2). On the other hand, some regions promote "coal-to-biomass" projects to extend the life span of fossil coal plants, thus benefiting from subsidies, as biomass is compatible with existing coal plants [74][75]. The latter practice could provide cost-efficiency in the near-term pledge pathway (2030), as one can fulfil the targets of renewable shares and decarbonisation with renewable biomass (e.g., wood pellets are regarded as a zero-carbon feedstock in many nations, even when they are actually processed from imported stemwoods). But it might hinder the transition to superior alternatives with higher cost- and land-use-effectiveness towards 2050, especially when dedicated bioenergy power plants will still be operating given life spans of up to 60 years [74].

Therefore, the challenge of competing bioenergy demand opens up a window for energy modelling studies, to consider the allocation of scarce sustainable bioenergy through sector-coupled energy system models.

# 5. Conclusion and Policy Implications

This study identified three challenges and opportunities of EU bioenergy by analysing bioenergy's historical national deployment, current policy support, and possible future roles in Europe. (1) On the supply side, the security of bioenergy supply is challenging for liquid biofuels and those countries with the highest per-capita bioenergy consumption in Europe. (2) The definition of "sustainable bioenergy" in modelling studies is sometimes inconsistent with how EU policies label it. (3) On the demand side, there are unique but competing uses for bioenergy without a clear long-term strategy in Europe. We conclude with three opportunities to tackle these challenges for future research. (1) Utilising the untapped bioenergy potential with low environmental impacts could improve supply security. (2) A clear and harmonised definition of "sustainable bioenergy" could better convey modelling results to policymaking. (3) Understanding where best to use limited sustainable bioenergy supply through sector-coupled energy system models can provide direction for a clearer EU bioenergy strategy towards 2050.

These conclusions have two-fold policy implications for both national and Europe-wide policymakers. First, more national policy should tackle the trilemma of biofuels supply security in consumption-leading countries – high import dependency, high bioenergy share, and the low subsidised level (especially in the Nordic region; see Figure 8). Apart from energy security regulations, stringent sustainability criteria or voluntary sustainable certificates should cover non-EU sourced biomass imports to prevent deforestation or environmental burdens in sourcing countries with lower sustainability standards. Second, EU policymakers and energy modellers should collaboratively close the gap of "sustainability assumptions" and explore the realistic long-term role of bioenergy. Through modelling competing usages and local endowments, a better understanding of the most strategical use of bioenergy per country or even per region could help EU policymakers develop a more coherent bioenergy strategy towards 2050.

## Data availability

The data for reproducing quantitative results and figures are available in the open repository of this paper: https://github.com/wwwuFei/Data-for-Challenges-and-Opportunities-of-EU-Bioenergy

## Acknowledgements

This project has received funding from the European Union's Horizon 2020 research and innovation programme under the Marie Sklodowska-Curie (MSC) grant agreement No. 847585, and from the SENTINEL project of the European Union's Horizon 2020 research and innovation programme under grant agreement No. 837089. The authors would like to thank Dr. Karin van Nispen for her professional language advice; Mr. Ziyan Gao from Shanghai Jiao Tong University and Mr. Chenxing Ji from TU Delft for helping with data processing and coding. The authors are grateful for the valuebale comments from anonymous reviewers.